\documentclass[aps,prb,twocolumn,a4paper,floatfix,showpacs,superscriptaddress]{revtex4}
\usepackage{graphicx}
\usepackage{hyperref}
\usepackage{color}
\usepackage{amsmath}
\usepackage[normalem]{ulem}
\usepackage{lineno}
\usepackage{multirow}
\usepackage{makecell}
\usepackage{float}
\usepackage{boondox-cal}

\begin{document}

\title{The underestimation of high pressure in DFT+$U$ simulation for the wide range cold-pressure of lanthanide metals}

\author{Yue-Chao Wang}
\affiliation{Laboratory of Computational Physics, Institute of Applied Physics and Computational
Mathematics, Beijing 100088, China}

\author{Bei-Lei Liu}
\affiliation{Laboratory of Computational Physics, Institute of Applied Physics and
Computational Mathematics, Beijing 100088, China}
\affiliation{School of Mathematical Sciences, Beijing Normal University, Beijing 100875, China}

\author{Yu Liu}
\email{liu\_yu@iapcm.ac.cn}
\affiliation{Laboratory of Computational Physics, Institute of Applied Physics and Computational Mathematics, Beijing 100088, China}

\author{Hai-Feng Liu}
\affiliation{Laboratory of Computational Physics, Institute of Applied Physics and
Computational Mathematics, Beijing 100088, China}

\author{Yan Bi}
\affiliation{Center for High Pressure Science and Technology
Advanced Research, Beijing 100094, China}
\affiliation{National Key Laboratory for Shock Wave and Detonation Physics,
Institute of Fluid Physics, CAEP, Mianyang 621900, China}

\author{Xing-Yu Gao}
\affiliation{Laboratory of Computational Physics, Institute of Applied Physics and Computational Mathematics, Beijing 100088, China}

\author{Jie Sheng}
\affiliation{Laboratory of Computational Physics, Institute of Applied Physics and Computational Mathematics, Beijing 100088, China}
\affiliation{Graduate School of China Academy of Engineering Physics, Beijing 100088, China}

\author{Hong-Zhou Song}
\affiliation{Laboratory of Computational Physics, Institute of Applied Physics and
Computational Mathematics, Beijing 100088, China}

\author{Ming-Feng Tian}
\affiliation{Laboratory of Computational Physics, Institute of Applied Physics and
Computational Mathematics, Beijing 100088, China}

\author{Hai-Feng Song}
\email{song\_haifeng@iapcm.ac.cn}
\affiliation{Laboratory of Computational Physics, Institute of Applied Physics and
Computational Mathematics, Beijing 100088, China}

\date{\today}

\begin{abstract}
Density functional theory plus $U$ (DFT+$U$) is one of the most efficient first-principles
methods to simulate the cold pressure properties of strongly-correlated materials.
However, the applicability of DFT+$U$ at ultra-high pressure is not sufficiently studied,
especially in the widely-used augmented schemes [such as projector augmented wave (PAW)
and linearized augmented plane wave (LAPW)].
This work has systematically investigated the performance of DFT+$U$ in PAW and LAPW at
the pressure up to several hundred GPa for the lanthanide metals, which is a typical
strongly-correlated series. We found DFT+$U$ simulation in PAW exhibits an unphysical
underestimation of pressure in high-pressure region, \textcolor{black}{which was not observed in
conventional DFT simulation.} By delicate analysis and comparison with
local-orbital-independent hybrid functional results, we have demonstrated that this unphysical
behavior is related to a normalization problem on the local density matrix caused by the overlap
of local orbitals in PAW under high pressure. Additionally, we observed a slight underestimating of pressure
in correlation correction methods in heavy lanthanides (Tm, Yb and Lu) in high-pressure region comparing with the DFT results
without the influence of local orbital overlap,
\textcolor{black}{and it might be related to the enhancement of binding between atoms in correlation
correction methods at high pressure.}
Our work reveals the
underestimation of high pressure in DFT+$U$ simulation, analyses two sources of this unusual
behavior and proposes their mechanism. Most importantly, our investigation highlights the
breakdown of DFT+$U$ for high pressure simulation in VASP package based on PAW framework.
\end{abstract}

\maketitle

\section{Introduction}
Equation of state (EOS) which describes the states and thermodynamic properties of
matter is of great interest in astrophysics, planetary physics, power engineering,
controlled thermonuclear fusion, impulse technologies, engineering, and several special
applications\cite{Bushman1992}. The cold pressure of EOS is usually reduced from shock wave
experiment and calculated from Thomas-Fermi model\cite{Feynman1949} with various
corrections\cite{McCarthy1965, Kirzhnits1975}. It is very necessary to validate the cold pressure
by rigorous theoretical models and precision
experimental data\cite{Liu2016, Young2016, Song2009, Song2007}.
Lanthanide materials as typical research objects for high-pressure physics and national defense,
its wide-range cold pressure is of especial importance\cite{Grasso2013, Samudrala2013}.
However, it is still difficult to obtain the high-quality experimental results under
extreme cold-pressure conditions. As far as we know, the highest static pressure achieved
in lanthanide materials is the recent works on Ho (282 GPa) and Nd (302 GPa),
and experimental works with static pressure over 250 GPa are rare\cite{Finnegan2021, Perreault2020}.
Thus, the simulation becomes further more important to obtain high-pressure data,
especially the high precision first-principles methods \cite{Sun2016,Tsuchiya2003,Zhang2017}.
However, the systematic research on the applicability of these conventional methods is still
lack of simulating the strong-correlated electronic materials, such as lanthanide metals.

Until now, the most successful and widely used first-principles methods are often referred to the
Kohn-Sham density functional theory (KS-DFT), and the correction methods based on
it\cite{Martin2004,Jain2016}. For the simulation of EOS at conventional conditions, many works have
proven the capability of KS-DFT on weakly correlated material\cite{Soderlind2018, Zhang2021}.
Among the implementation techniques of DFT, the augmented methods, such as linearized augmented
plane wave (LAPW) and projector augmented wave (PAW) are of most widely
usage \cite{Weinert1981,Wimmer1981,Vanderbilt1990,Blochl1994}. LAPW and PAW are now implemented
in many famous software packages, e.g. WIEN2k \cite{Blaha2020}, Elk\cite{Elk},
VASP\cite{Kresse1999}, Quantum-Espresso\cite{Giannozzi2020}, GPAW\cite{Enkovaara2010}
and ABINIT\cite{Gonze2020}. LAPW and PAW methods are the representatives of high accuracy
and high effectiveness in the conventional implementation of KS-DFT
respectively\cite{Holzwarth1997}. Another advantage of the augmented methods is that the
definition of atomic orbitals is straightforward in these methods. It makes the implementation
of correction methods based on local orbitals much easier in the
scheme of LAPW and PAW\cite{Kresse1999,Shick1999}.

For strongly correlated materials, such as lanthanides, the KS-DFT often fails,
and correction methods are needed. DFT+$U$ is one of the most effective correction methods,
which is based on local orbitals\cite{Anisimov1997,Liechtenstein1995}.
DFT+$U$ has been successfully used in the simulation of crystal structures, magnetic order,
charge order and many other properties related to the total
energy\cite{Dompablo2011,Himmetoglu2014}. There are other correction methods based on local
orbitals, such as DFT plus dynamic mean-field theory (DMFT) and Gutzwiller projected variational
wave function\cite{Anisimov1997,Deng2008}. These local-orbital-based corrections can make
delicate or effective corrections for the strongly-correlated part (like 4$f$-electron in
lanthanides) \cite{Anisimov2010, Avella2012}. Another class of correction methods is
non-local-orbital based, such as hybrid functionals. This class of methods makes no bias on
the electrons and often with less artificial factor (like the definition of local orbitals)
\cite{Casadei2016,Perdew1996}. However, the computational cost of hybrid functional is at least
one order larger than DFT and DFT+$U$\cite{Jiang2015}.
\textcolor{black}{The simulation of the wide range EOS of the complex materials usually requires
 tens or hundreds of total energy data points, which is beyond the capability of many
 sophisticated correction methods.}
For the calculation of wide range EOS, the most attractive candidate (if not only) is
the DFT+$U$ method, which applies a correction based on static mean-field approximation of
the onsite Hubbard model and has a similar computational cost as
KS-DFT\cite{Rahm2009,Modak2013,Steneteg2012,Hsu2011}.

The effectiveness of DFT+$U$ with PAW and LAPW methods has been just proved in conventional
conditions. But there have already been many works considering the strongly correlated effect
with DFT+$U$ performed for high-pressure condition, especially in
PAW\cite{Dompablo2011,Rollmann2005,Geng2007,Song2009-2,Zhang2010,Wen2013,Zhou2014}. Many analyses
and conclusions are based on these simulations, however, the applicability of DFT+$U$ for high
pressure is still not clear. In DFT+$U$ methods, many factors can affect its performance, like the
double-counting term, interaction strength $U$ and $J$, and the local orbitals constructed from DFT
calculation. Especially for the local orbitals defined in low pressure conditions, the overlap
problem at high pressure makes the reliability of local orbitals more suspicious than other factors.
In an earlier work, the definition and overlap of the local orbital in VASP has shown a different
performance in the Mn dioxides comparing with WIEN2k\cite{Wang2016}. This problem may become
serious when the distance between atoms is shortened to 80$\%$ at pressure over 100 GPa. Although
software such as VASP has already warned users about the risk of the overlap, the requirement of
the high-pressure simulation still makes researchers to take the risk. Thus, systematic research
of the reliability of DFT+$U$ in high pressure is necessary.

In this work, we investigate the performance of DFT+$U$ in lanthanide metals at high pressure
from two aspects. Firstly, we compare the DFT+$U$ results from VASP and WIEN2k at the
same $U$ and $J$ values at normal pressure and high pressure over 300 GPa, together with a set
of results of hybrid functional (HSE06) as a benchmark. From our systematic investigation of the
total energy, pressure and electronic structures of lanthanide metals, we find over unphysical
softening results given by the DFT+$U$ with PAW methods. Through our analysis, this over softening
connects with the normalization error of local orbitals led by the overlap. We also find the
results from higher reliable correction methods like hybrid functional and DFT+$U$ based on LAPW
show a trend of underestimating the pressure in high-pressure region comparing with the DFT. This phenomenon implies besides
the technique problem, there may be some physical mechanism.

The paper is organized as follows. In Sec. II we introduce the methods and parameters used
in this paper. In Sec. III, the total energy, pressure and electronic structure results of
lanthanide metals at different simulation conditions are exhibited. In Sec. IV we analyze the
abnormal underestimation of pressure in DFT+$U$ from the overlap of local orbitals view,
and discuss the results in Sec. III from the normalization error led by the overlap and an underestimating of pressure
caused by the mechanism of correction methods. In Sec. V we close the paper with a summary of the
main finds of this work and some general remarks. In the Appendix, the detailed procedure of
definition of local orbitals in LAPW and PAW are shown.

\section{Method}

\subsection{Simulation methods}

All 15 kinds of lanthanide metals investigated in this work are the face centered cubic (FCC)
structure, which is a conventional close-packed structure for lanthanide metals.
Since the main purpose of this work is to reveal the capability of DFT+$U$ method to simulate
metals with 4$f$ electrons under high-pressure, the other probable structures are not considered
in this work. To include both low and ultra-high pressure, the volumes we considered are from
9 ${\rm A^3}$ to 28 ${\rm A^3}$. PAW and LAPW simulations are performed in VASP\cite{Kresse1999}
and WIEN2k\cite{Blaha2020} packages respectively, and we specially refer PAW (LAPW) results to VASP
(WIEN2k) simulations in this work. The exchange-correlation functional used for all calculations
is generalized gradient approximation (GGA) of Perdew-Burke-Ernzerhof formation
(PBE)\cite{Perdew1996-2}. A grid of $17\times17\times17$ is used for the k-spacing sampling.
The pseudo-potentials used for PAW simulation are recommended by VASP,
\textcolor{black}{which means the 5$s$, 5$p$, 5$d$, 4$f$ and 6$s$ are treated as valence for all lanthanides.}
And the cutoff energy of plane wave is 600 eV, which is large enough for calculation of energy
and force for all conditions in this work. The augmentation radius of all 4$f$ elements in LAPW
calculation is 1.11 A (2.1 bohr.), and the \textcolor{black}{$R_{\rm kmax}$}. The ``+$U$'' correction takes the
isotropic form to reduce the number of parameters. The $U_{\rm eff}$ (or the $U-J$ in some context)
is 5.0 eV for all lanthanide elements, and the double-counting term is ``fully localized limit''
(FLL), which is the rational option for many DFT+$U$ works\cite{Jiang2009,Moree2018}.
For comparison, hybrid functional calculations using HSE (Heyd-Scuseria-Ernzerhof) formation are
performed in VASP\cite{Heyd2003}. Since spin-polarization can drastically change the electronic
states between different methods and makes the comparison much complicated, no spin-polarization is
considered in this work. For the calculation of pressure, each energy-volume curve is fitted by
Birch-Murnaghan equation with 50 sampling points in the range from 9 ${\rm A^3}$ to 28 ${\rm A^3}$.

\subsection{Projection of local orbits}
In the ``+$U$'' correction, besides the correlation strength $U_{\rm eff}$ and double-counting
term, one of the most important factors is the definition of local Hilbert space which the model
Hamiltonian is worked on\cite{Ryee2018}. From the technical point of view, a set of local orbitals
is often used to define this local subspace. By acting the projection operator $\hat{P}(\mathbf{r},
\mathbf{r}^\prime)$ composed by the local orbitals on the whole Hilbert space, the local density
matrix can be obtained\cite{Wang2016}. There are three kinds of most common formations of the
projection operator: spherical augmented schemes, atomic orbitals and special functions
(e.g. Wannier function)\cite{Shick1999, ORegan2010, Bengone2000, Cococcioni2005}.
As the simulation tools of materials based on augmented basis sets (e.g. VASP and WIEN2k) are widely
used, many works of ``+$U$'' correction are carried out with spherical augmented projection
operator. The details of projection operators in PAW and LAPW and how they act on the single
electron orbitals to obtain local density matrix are described in the appendix part.

\section{Results}

\subsection{Pressure in different methods}
\begin{figure}[htbp]
\centering
\includegraphics[width=0.48\textwidth]{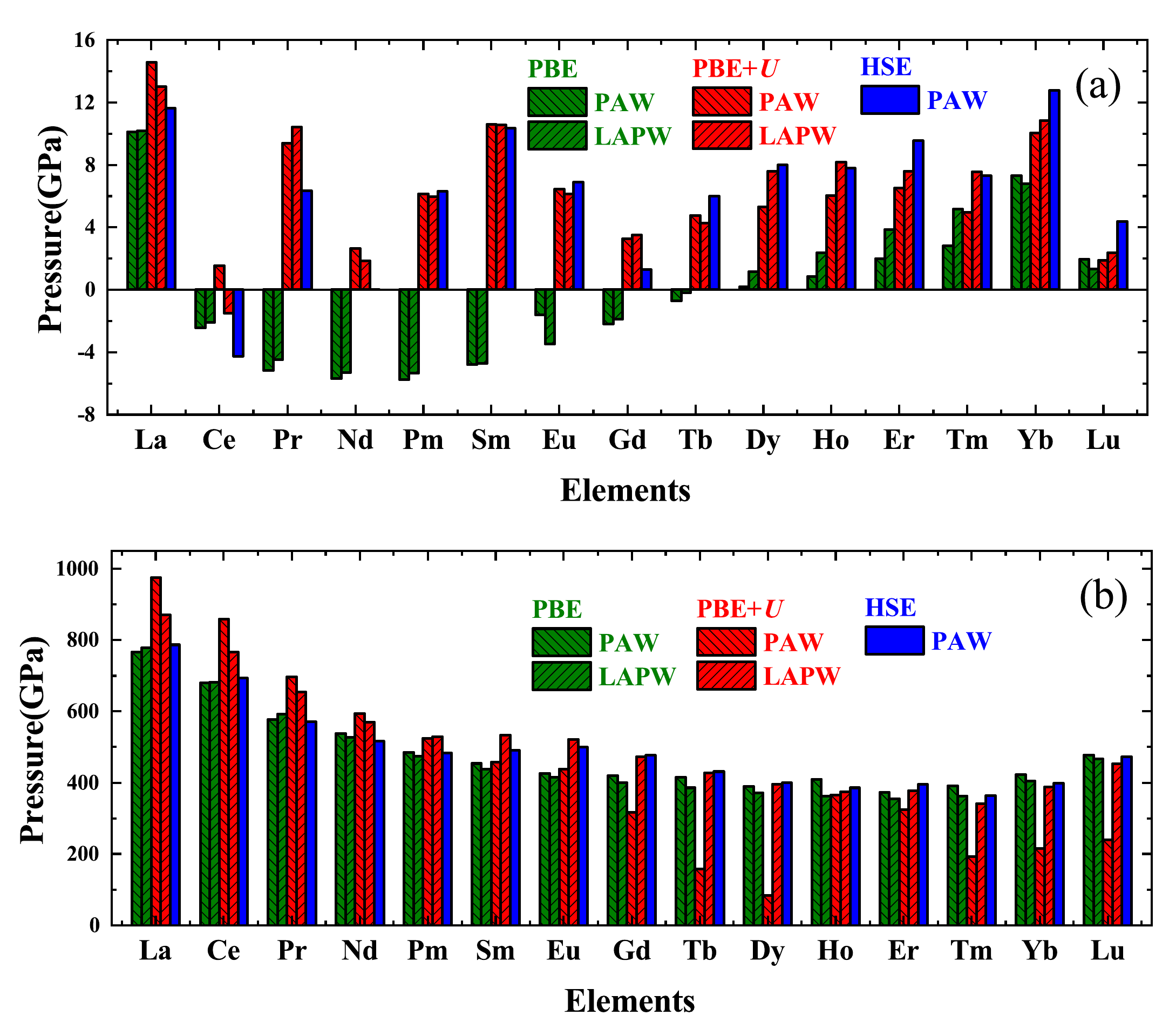}
\caption{\label{fig:stat-pres}The pressure calculated with different methods at large volume (28 $\rm A^3$, panel a) and small volume (9 $\rm A^3$, panel b) of lanthanide metals with FCC structure. The green bars stand for the results from DFT (PBE functional), the red bars stand for the results from DFT+$U$ ($U_{\rm eff}$= 5.0 eV), and the blue bars stand for the results from HSE06. The VASP and Wien2k results are mark by shadow with backslash and slash respectively. }
\end{figure}

The results of lanthanide metals' pressure at large (upper panel, 28 $\rm A^3$) and small
(lower panel, 9 $\rm A^3$) volume in different simulation conditions are shown in Fig.\ref{fig:stat-pres}.
The simulation condition we considered includes DFT (PBE, green bar) and ``+$U$'' correction
(red bar) in PAW and LAPW methods, and the HSE06 hybrid functional (blue bar) based on PAW.
HSE06 is used as a benchmark, considering hybrid functional is not directly dependent on the local
orbitals as ``+$U$'' correction, and the level of correction of correlation effect is not lower than
the ``+$U$'' correction. Two important features are exhibited in Fig.\ref{fig:stat-pres}.
Firstly, an abnormal high-pressure softening of the results from DFT+$U$ in PAW can be observed.
In high pressure conditions, the pressure of some heavy lanthanide metals is underestimated by
50$\%$ to 80$\%$ in DFT+$U$ simulation from PAW comparing with that from LAPW. Meanwhile,
in low pressure conditions, the DFT+$U$ results between PAW and LAPW are similar.
Secondly, the pressure calculated by DFT is higher than the correlation correction methods
(both ``+$U$'' and HSE06) in some heavy lanthanide metals (Tm, Yb and Lu). It is contrary to the
general experience of these correction methods that ``+$U$'' and hybrid functional often give a
higher pressure comparing to DFT at the same volume, as can be observed in our results of
low-pressure condition\cite{Wang2016,Perdew1996,Jiang2009,Moree2018}. For the first feature
mentioned above, it can be found that the PAW pseudo-potential is not the reason of the abnormal
softening of pressure from DFT+$U$ in PAW, since both in high and low pressure the DFT results from
LAPW and PAW are similar. The ``+$U$'' correction also should not be accused for these abnormal
results, since HSE06 and ``+$U$'' correction in LAPW give similar results in both low and high
pressure. The answer to the unreasonable results of ``+$U$'' correction in PAW under high pressure
seems only related to the implementation of ``+$U$'' correction in PAW. The main difference of
implementation ``+$U$'' in PAW and LAPW is that the overlap between augmented sphere in PAW cannot
be avoided, since the radius of the sphere is fixed when the pseudo-potential is created in PAW.
Other results connecting with the abnormal behavior of PAW in high-pressure condition is shown in
the following, and a detailed analysis is also presented. For second feature that the pressure
described by DFT is higher than that of correlation correction in high-pressure conditions, it may
relate to the theorical mechanism behind correlation correction methods other than the
implementation problem, since the phenomenon exhibits in both HSE06 and ``+$U$'' simulation. We
give our interpretation of the phenomenon in the discussion.

\subsection{Energy, Pressure and Modulus versus volume}

\begin{figure}[htbp]
\centering
\includegraphics[width=0.48\textwidth]{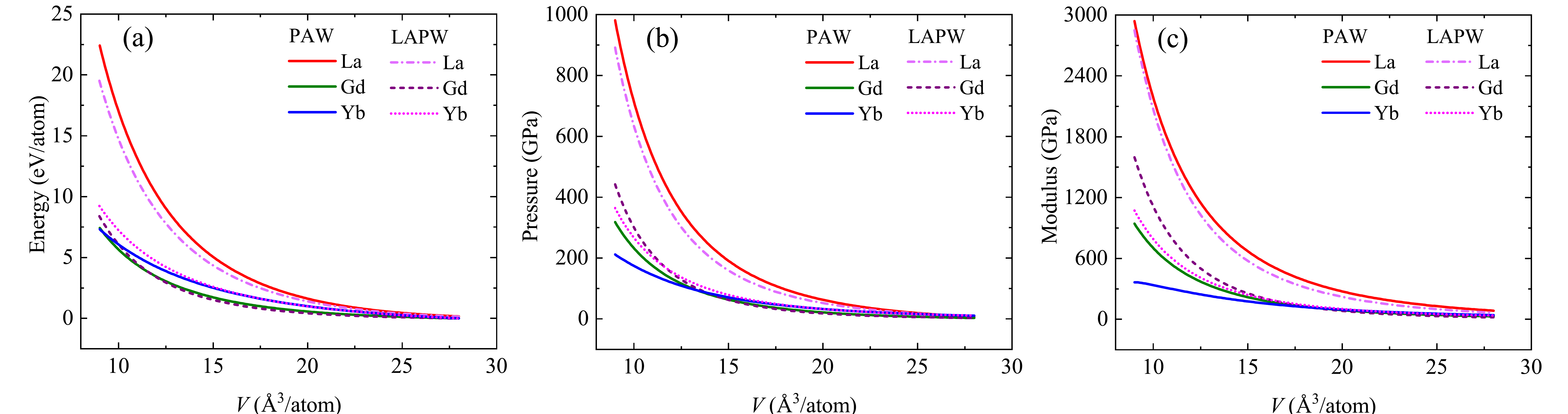}
\caption{\label{fig:EvsV}
The results of total energy (a), pressure (b) and modulus (c) versus volume from DFT+$U$ in 6
typical lanthanide metals. Lower occupied systems (La) are red solid line (PAW) and violet
dot-dashed line (LAPW). Half occupied systems (Gd) are green solid line (PAW) and purple dashed
line (LAPW). Full occupied systems (Yb) are blue solid line (PAW) and pink dotted line (LAPW).
The zero point of total energy is set as the energy of the smallest volume (9 $\rm A^3$).}
\end{figure}

In this subsection, the ``+$U$'' results of total energy, pressure and modulus changing versus
volume (simplified to E-V, P-V and B-V in the work respectively) are displayed in PAW and LAPW.
The E-V, P-V and B-V results are calculated in the range from 9 $\rm A^3$ to 26 $\rm A^3$. We have
checked the pressure from the Hellmann-Feynman (reading from the software report) and from energy
difference, and the results are the same. It should be mentioned here, the zero point of the total
energy of each system is set as the energy of 9 $\rm A^3$. For the simulation in this subsection,
we display the results of La, Gd and Yb, since these three lanthanide metals represent three kinds
of 4$f$-occupation conditions, the empty occupation La, half occupation Gd and full occupation Yb.
The Fig.\ref{fig:EvsV} shows the E-V, P-V and B-V results of La, Gd, and Yb. For La, the E-V results from PAW
always change fast comparing to LAPW with volume decreasing. This is obvious in the P-V results,
for pressure can be seen as the first derivative of energy with respect to volume. In the P-V
results of La, the pressure of PAW is always large than that of LAPW at every volume point.
However, it can be observed that the difference of pressure between PAW and LAPW is decreasing with
the decreasing of volume. In the B-V results (the second derivative of energy with respect to volume),
the difference between PAW and LAPW is almost vanishing at the small volume. For half occupation
system Gd, with the decrease of volume, the total energy of PAW increases faster than LAPW below
100 GPa. At higher pressure, results of LAPW increase faster. In the P-V results, curves of PAW and
LAPW cross at about 100 GPa in Gd. These crossings happen at a much larger volume in the B-V results
of Gd. For the nearly fully occupied Yb, the variation of energy versus volume in PAW and LAPW is
similar in low pressure. In P-V results, the pressure of LAPW is larger than PAW results at every
volume point, and at the volume of 9 $\rm A^3$, the difference can be as large as 150 to 200 GPa,
which has a relative difference larger than 50$\%$. For the B-V results, PAW gives smaller values
comparing with LAPW, and an abnormal phenomenon is observed that modulus from PAW starts to
decrease with the decreasing of volume at high pressure.

\subsection{Electronic structures}

\begin{figure}[htbp]
\centering
\includegraphics[width=0.48\textwidth]{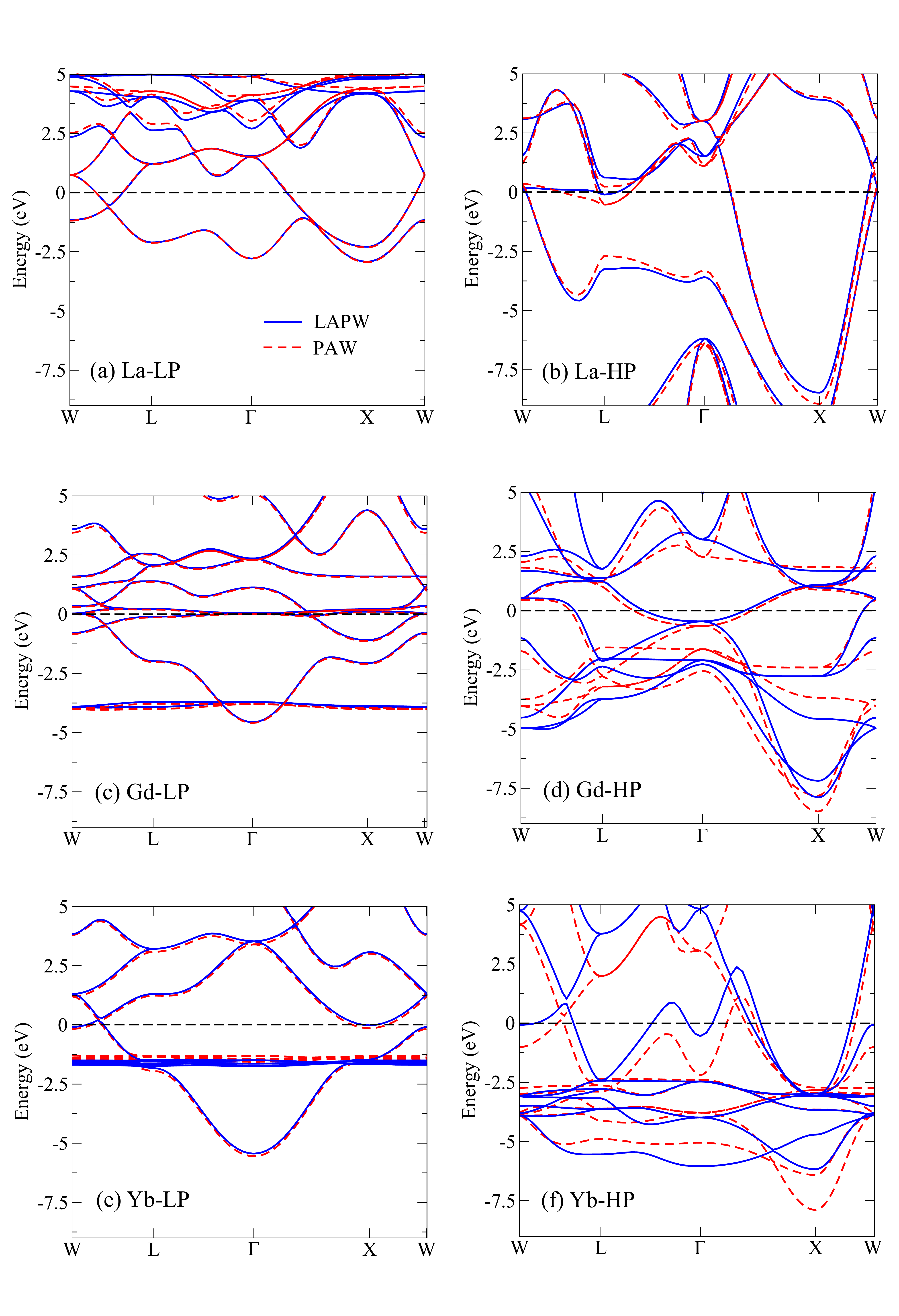}
\caption{\label{fig:bands}The electronic band structure calculated by DFT+$U$ in VASP (red dashed line) and WIEN2k (blue solid line) of La (a,b), Gd (c,d) and Yb (e,f) at low pressure (LP) and high pressure (HP).}
\end{figure}

The ``+$U$'' correction of electronic structures at low and high pressure is investigated in this
subsection. We also choose La, Gd and Yb as representatives for empty, half-filling and full
occupied $f$-shell. The band structures calculated by the ``+$U$'' correction in PAW and LAPW are
shown in Fig.\ref{fig:bands}. At low pressure, it can be observed that the electronic structures
from PAW and LAPW are similar, especially the itinerant bands below the Fermi level. For the local
$f$-bands, the results of PAW and LAPW show the difference about 0.1 eV. The PAW results give lower
$f$-bands below the Fermi level than LAPW results. At high pressure, the difference of electronic
structures calculated from PAW and LAPW becomes larger. Besides all bands becoming more dispersed,
PAW still makes the $f$-bands far away from the Fermi level comparing to LAPW, but the detailed
features become much more complicated. However, it seems no qualitative difference between
PAW and LAPW electronic structures at high pressure.

\section{Discussion}

\subsection{Total Energy and Occupation}
To understand the abnormal pressure feature shown in the results of ``+$U$'' in the PAW method,
we should analyze how ``+$U$'' correction makes an influence on the total energy. It is because the
pressure reflects the changing of total energy versus volume. The total energy of DFT+$U$ is
\begin{align}
E_{{\mathrm{DFT+}}U}&\left[\rho\left(\mathbf{r}\right),\left\{n\right\}\right]\nonumber\\
&=E_{\rm DFT}\left[\rho(\mathbf{r})\right]+E_U\left[{n}\right]-E_{\rm dc}\left[{n}\right].
\end{align}
In this work,
\textcolor{black}{all formations of ``$+U$'' correction are shown on a diagonalized
local density matrix. On one hand, it can make the analysis more clear. On the other hand, for the
examples in this work, the diagonal part gives the major contribution to the results.
The double counting of FLL form is adopted, which is}
\begin{align}
E_{\rm dc}^{\rm FLL}=\frac{U}{2}N\left(N-1\right)
-\sum_{\sigma}{\frac{J}{2}N_\sigma\left(N_\sigma-1\right)}.
\end{align}
Without the anisotropic effect, the total energy correction in ``+$U$'' is
\begin{align}
\Delta E_{{\rm DFT}+U} =\frac{U_{\rm eff}}{2}\sum_{m,\sigma}{[n_m^\sigma-({n_m^\sigma})^2]}.
\end{align}
Since in most of the ``+$U$'' works, the $U_{\rm eff}$ is constant, $n_m^\sigma$ the occupation of
local orbitals, becomes the key factor that affects the total energy.

The $n$ should have a value between 0 and 1 from the definition of the occupation of local orbitals.
It is because $n_{mm}$ is
\begin{align}
\label{eq:projnn}
n_{mm}=\sum_{i\in occ}{\left\langle m\middle|\psi_i\right\rangle\left\langle\psi_i\middle|
m\right\rangle}.
\end{align}
$\psi_i$ is KS orbital, and the summation is over all occupied orbitals.
For the completeness of the KS orbitals, all the KS orbitals can construct a unit operator
\begin{align}
\hat{I}=\sum_{i}^{\infty}\left|\psi_i\right\rangle\left\langle\psi_i\right|.
\end{align}
Thus, it can be written out like
\begin{align}
\nonumber
\left\langle m\middle| m\right\rangle &= \sum_{i}^{\infty}{\left\langle m\middle|\psi_i\right\rangle\left\langle\psi_i\middle| m\right\rangle}
\\
&=\sum_{i\in o c c}{\left\langle m\middle|\psi_i\right\rangle\left\langle\psi_i\middle| m\right\rangle}+\sum_{i\notin o c c}{\left\langle m\middle|\psi_i\right\rangle\left\langle\psi_i\middle| m\right\rangle}.
\end{align}
So,
\begin{align}
n_{mm}=\left\langle m\middle| m\right\rangle-\sum_{i \notin occ}{\left\langle m\middle|\psi_i\right\rangle\left\langle\psi_i\middle| m\right\rangle}
\end{align}
Because $\sum_{i\notin occ}{\left\langle m\middle|\psi_i\right\rangle\left\langle\psi_i\middle| m\right\rangle}\geq 0$, $n_{mm}$ should little than $\left\langle m\middle| m\right\rangle$. If $\left|m\right\rangle$ is a well-defined atomic orbital, it should satisfy the normalization condition that $\left\langle m\middle| m\right\rangle=1$, and we can make a conclusion that $n_{mm}\leq 1.0$.

For the n with value in range 0 to 1 and $U_{\rm eff}$ $>$ 0.0 (electrons repulse with each other),
the ``+$U$'' correction with FLL double counting term gives an energy correction larger than 0.0.
This correction becomes 0 only if n equals 1 or 0. This feature of ``+$U$'' correction has been
reported in previous works. Another well-known feature is that lower orbital energy is obtained
with occupation larger than 0.5\cite{Ylvisaker}.

\subsection{Occupation problem in PAW}

In our PAW simulations, the occupation of some local density matrix elements exceeds 1.0,
and we investigate the origin and influence of the phenomenon through the variation of local
density matrix with the pressure (or volume). $\Delta N_{>1}$ is defined as the summation of all
local orbitals which is more than 1.0, and has the form
\begin{align}
\Delta N_{>1} = \sum_{m = -3}^{3}
\begin{cases}
n_{mm}-1.0 & n_{mm}>1\\
0.0 & n_{mm}\leq 1.
\end{cases}
\end{align}
$n_{mm}$ are the diagonal elements of the local density matrix,
and the summation of m from -3 to 3 means all 7 orbitals in 4$f$ shell
(spin polarization is not considered).

\begin{figure}[htbp]
\centering
\includegraphics[width=0.48\textwidth]{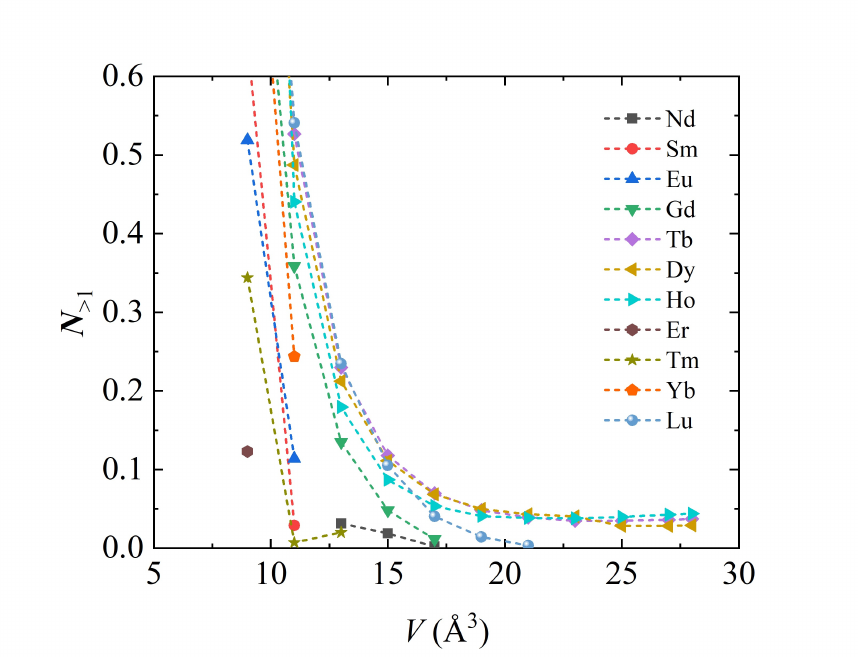}
\caption{\label{fig:occ}The over 1.0 part of the local orbital occupation of the lanthanide
metals versus volume in the VASP DFT+$U$ calculation. The systems without any local orbital
occupation over 1.0 (La, Ce, Pr) are not shown in the figure.}
\end{figure}

As shown in Fig.\ref{fig:occ}, except for some elements with lower occupation (like, La, Ce, Pr),
the occupation of most lanthanide metals can show obvious exceeding over 1.0. The systems with
occupation over 1.0 can be divided into two kinds. One is that the occupation of some orbitals is
over 1.0 even at low pressure (Tb, Dy, Ho). These systems correspond to the first dipping of
pressure in serial of lanthanide metals in high pressure results. In the other kind occupation only
exceeds 1.0 when the pressure is high enough. No matter which kind of behavior, as we have discussed
above, the occupation of local orbital exceeding 1.0 happens only in two conditions:
The normalization of KS orbitals is not satisfied, or the modular square of local orbitals has
problem. As the DFT results of PAW and LAPW are similar, the KS-orbital should be normalized
correctly. We also have tested the modular square of KS-orbitals dominated by $f$-feature, and the
normalization can be achieved at $10^-8$ level. To analyse the local orbitals, we split the modulus
square of PAW orbitals into several parts. The modular square of the PAW orbitals writes

\begin{align}
\nonumber
\left\langle\psi_i\middle|\psi_i\right\rangle&=\left\langle{\widetilde{\psi}}_i\middle|{\widetilde{\psi}}_i\right\rangle
\\
&+\sum_{ss^\prime}{\left\langle{\widetilde{\psi}}_i\middle|{\widetilde{p}}_s\right\rangle\left(\left\langle\phi_s\middle|\phi_{s^\prime}\right\rangle-\left\langle{\widetilde{\phi}}_s\middle|{\widetilde{\phi}}_{s^\prime}\right\rangle\right)\left\langle{\widetilde{p}}_{s^\prime}\middle|{\widetilde{\psi}}_i\right\rangle}.
\end{align}
$\left|\phi_s\right\rangle$ and $\left|{\widetilde{\phi}}_s\right\rangle$ are the all-electron partial orbital and pseudo partial orbital of the same atom respectively. $\left|\psi_i \right\rangle$ and $\left|{\widetilde{\psi}}_i \right\rangle$ are all-electron orbital and pseudo-orbital and these two terms are connected by a projector $\left|{\widetilde{p}}_s\right\rangle$. In usual case, $\left|\phi_s\right\rangle$, $\left|{\widetilde{\phi}}_s\right\rangle$ and $\left|{\widetilde{p}}_s\right\rangle$ is fixed as a certain pseudo-potential is used, while $\left|\psi_i \right\rangle$ and $\left|{\widetilde{\psi}}_i \right\rangle$ are renewed at each self-consistent cycle. The subscript s stands for a collection of orbital momentum l and m, and the site of atom. Here we split it into three parts:
integral of pseudo-orbital on plane wave
\begin{align}
{\widetilde{\mathcal{S}}}_{i,pw}=
\left\langle{\widetilde{\psi}}_i\middle|{\widetilde{\psi}}_i\right\rangle,
\end{align}
integral of pseudo atomic orbital
\begin{align}
\nonumber
{\widetilde{\mathcal{S}}}_{i,atom}&=\sum_{l}{\widetilde{\mathcal{s}}}_{i,atom}^{lm}
\\
{\widetilde{\mathcal{s}}}_{i,atom}^{lm}&=\sum_{s\in l m;s^\prime\in l m}
{\left\langle{\widetilde{\psi}}_i\middle|
{\widetilde{p}}_s\right\rangle\left\langle
{\widetilde{\phi}}_s\middle|{\widetilde{\phi}}_
{s^\prime}\right\rangle\left\langle{\widetilde{p}}_
{s^\prime}\middle|{\widetilde{\psi}}_i\right\rangle}.
\end{align}
and integral of all-electron atomic orbital
\begin{align}
\nonumber
\mathcal{S}_{i,atom}&=\sum_{lm}\mathcal{s}_{i,atom}^{lm}
\\
\mathcal{s}_{i,atom}^{lm}&=\sum_{s\in l m;s^\prime\in l m}{\left\langle{\widetilde{\psi}}_i
\middle|{\widetilde{p}}_s\right\rangle\left\langle\phi_s\middle|
\phi_{s^\prime}\right\rangle\left\langle{\widetilde{p}}_{s^\prime}\middle|{\widetilde{\psi}}_i
\right\rangle}.
\end{align}
With this splitting, the integral of PAW orbital can be written as
\begin{align}
\left\langle\psi_i\middle|\psi_i\right\rangle=
{\widetilde{\mathcal{S}}}_{i,pw}+\mathcal{S}_{i,atom}-{\widetilde{\mathcal{S}}}_{i,atom}.
\end{align}
Considering the definition of the diagonal element of local density matrix $n_{mm}$,
\begin{align}
n_{mm}=\sum_{i\in o c c}\mathcal{s}_{i,atom}^{lm}.
\end{align}
If $n_{mm}$ is fully contributed by one occupied KS-orbital
$\psi_i$, $n_{mm}=\mathcal{s}_{i,atom}^{lm}$. The interval part of the integral is
${\widetilde{\mathcal{S}}}_{i,pw}-{\widetilde{\mathcal{S}}}_{i,atom}$, and the atom part of the
integral is $\mathcal{S}_{i,atom}$. The interval and atom part of the integral stand for the
probability to find the electron on orbital in the interval and augmented region respectively.

\textcolor{black}{As we have shown, the all-electron atomic orbital has
a close relation to the local orbital.} And we can investigate the
normalization of local orbital by the modular square of all-electron atomic orbital. From the
physical meaning of orbital, $\mathcal{S}_{i,atom}$ should be less than 1.0, otherwise there is
something abnormal happened to the local orbital. We analyze this problem at high pressure in PAW
method in Yb as an example, since the abnormal behavior in high-pressure simulation is obvious in
the heavy lanthanides. \textcolor{black}{It should be clarified that the occupation defined in
Eq.\ref{eq:projnn} is not the exact case in augmented methods. However, if a case with only one
type of radial orbital for each shell is considered, the occupation defined in Eq.\ref{eq:projnn}
could be equivalent with the augmented methods. }

\textcolor{black}{Here we consider a simplified situation, where only one kind of radial part $|R_l\rangle$ is used for the shell $l$ in each site. In this way,
it is easier to compare the difference between occupation matrix elements obtained from augmented
methods and the ones from projecting onto a set of local orbital ${|m\rangle}$, as shown in Eq.\ref{eq:projnn}
The occupation matrix elements in augmented methods become,
\begin{align}
n_{mm'}^{\rm aug}=\sum_{i\in occ.}{\langle{\widetilde\psi}_i}|{\widetilde{p}_m\rangle}
{\langle\widetilde{p}_{m'}}|{\widetilde\psi_i\rangle}{\langle{R_l}|{R_{l}}\rangle}.
\end{align}
$n_{mm'}^{\rm aug}$ stands for an occupation matrix element in PAW, and a general form of it is provided in appendix. The occupation matrix elements obtained by projecting onto local orbitals are,
\begin{align}
n_{mm'}^{\rm proj}=\sum_{i\in occ.}{\langle{\widetilde\psi}_i}|{\widetilde{p}_m\rangle}
{\langle\widetilde{p}_{m'}}|{\widetilde\psi_i\rangle}|{\langle{R_l}|{R_{l}}\rangle}|^2.
\end{align}
$n_{mm'}^{\rm proj}$ is the occupation matrix element from the procedure defined in Eq.\ref{eq:projnn}, and the radial part of local orbital $|R_l\rangle$ takes the same function as the all-partial orbital in PAW. It can be clearly observed that if the radial parts of the shell has the same shape, and the
local orbital is well normalized, $n_{mm'}^{\rm aug}=n_{mm'}^{\rm proj}$.
}

In the table \ref{tab:int}, we show the different parts of orbital integral of Yb at low and high
pressure at $\Gamma$ point of reciprocal space. As in the ``+$U$'' correction of Yb, almost all
$f$-feature bands are below the Fermi energy, the $\mathcal{S}_{i,atom}$ and
${\widetilde{\mathcal{S}}}_{i,atom}$ are almost contributed by $\mathcal{s}_{i,atom}^{fm}$, and the
7 orbitals listed in the table are the ones with obvious $f$-feature. At low pressure, the integral
results perform normally. A finite positive value shows for every part, and the summation of all
parts equals 1.0. At high pressure, though the summation of all parts is still 1.0, there shows
some abnormal features. The part of the augmented region is over 1.0, and the interval region has
negative values. These features are consistent with the occupation of local orbitals over 1.0 at
high pressure and imply a normalization problem of the local orbital.

\begin{figure}[htbp]
\centering
\includegraphics[width=0.48\textwidth]{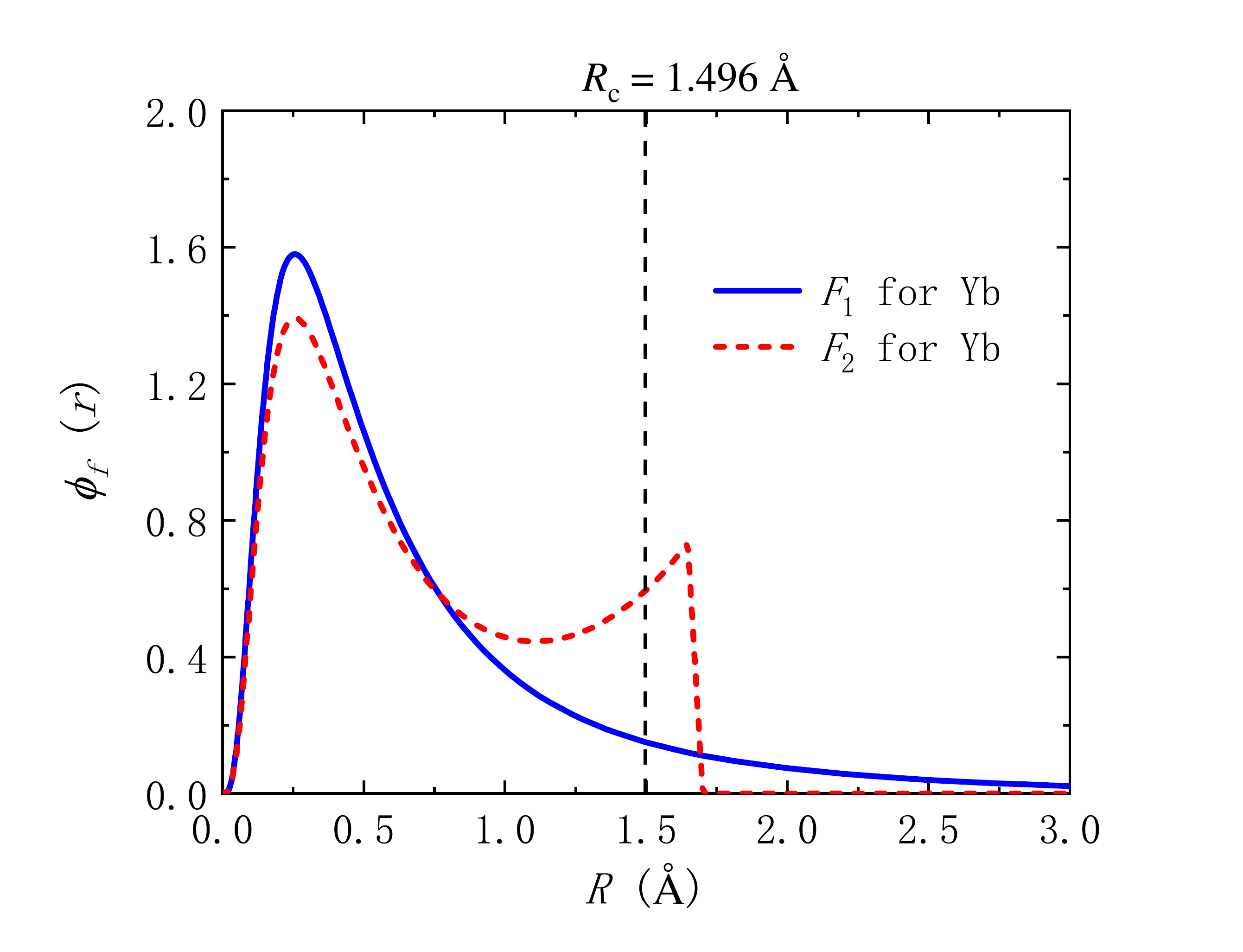}
\caption{\label{fig:fig5}Two prefixed the radial $f$ orbitals ($F_1$ and $F_2$) of
all-electron in the Yb PAW pseudo-potential of VASP (blue solid line and red dashed line). $R_{\rm c}$
stands for the augmented range of PAW, which is a range of projection operator in DFT+$U$.}
\end{figure}

The drastic changes of $\mathcal{S}_{i,atom}$ and ${\widetilde{\mathcal{S}}}_{i,atom}$ between
low-pressure and high-pressure conditions implies a variation of the wave function radial part with
pressure in the augmented region. In VASP, there are two predefined radial functions forming the
basis to describe the f-shell. We show these radial functions in Fig.\ref{fig:fig5}. The blue solid line
represents a radial function with typical 4$f$-feature (localized around core, one peak and
decaying to zero at distance), and we labeled it by $F_1$. The red dashed line represents a radial
function more divergent, we labeled it by $F_2$. The black vertical dashed line labeled by $R_{\rm c}$
stands for the radius of the augmented region. The sudden drop of $F_2$ is a truncation, and since
it is out of the $R_{\rm c}$, we believe it doesn't affect the results directly.

\begin{table}[htbp]
\caption{\label{tab:int}The components of the module square integral of KS orbitals with
4$f$ feature at Gamma point in Yb metal in low pressure (LP) and high pressure (HP) conditions}
\begin{tabular*}{0.45\textwidth}{ccccccc}
\hline\hline
\multirow{2}{*}{No.}& \multicolumn{2}{c}{$\mathcal{S}_{i,atom}$}& \multicolumn{2}{c}{\quad${\widetilde{\mathcal{S}}}_{i,atom}$}& \multicolumn{2}{c}{\qquad${\widetilde{\mathcal{S}}}_{i,pw}-{\widetilde{\mathcal{S}}}_{i,atom}$} \cr
    &LP& HP&\quad LP& HP&\quad LP& HP
\\
\hline
1           &0.9803  &1.3998  &\quad0.0601  &0.6521  &\quad0.0197  &-0.3998\\
2           &0.9904  &1.0474  &\quad0.0482  &0.1359  &\quad0.0096  &-0.0508\\
3           &0.9904  &1.0474  &\quad0.0482  &0.1359  &\quad0.0096  &-0.0508\\
4           &0.9904  &1.0474  &\quad0.0482  &0.1359  &\quad0.0096  &-0.0508\\
5           &0.9991  &1.0114  &\quad0.0345  &0.0199  &\quad0.0009  &-0.0114\\
6           &0.9991  &1.0114  &\quad0.0345  &0.0199  &\quad0.0009  &-0.0114\\
7           &0.9991  &1.0114  &\quad0.0345  &0.0199  &\quad0.0009  &-0.0114\\
\hline\hline
\end{tabular*}

\end{table}

\begin{figure}[htbp]
\centering
\includegraphics[width=0.48\textwidth]{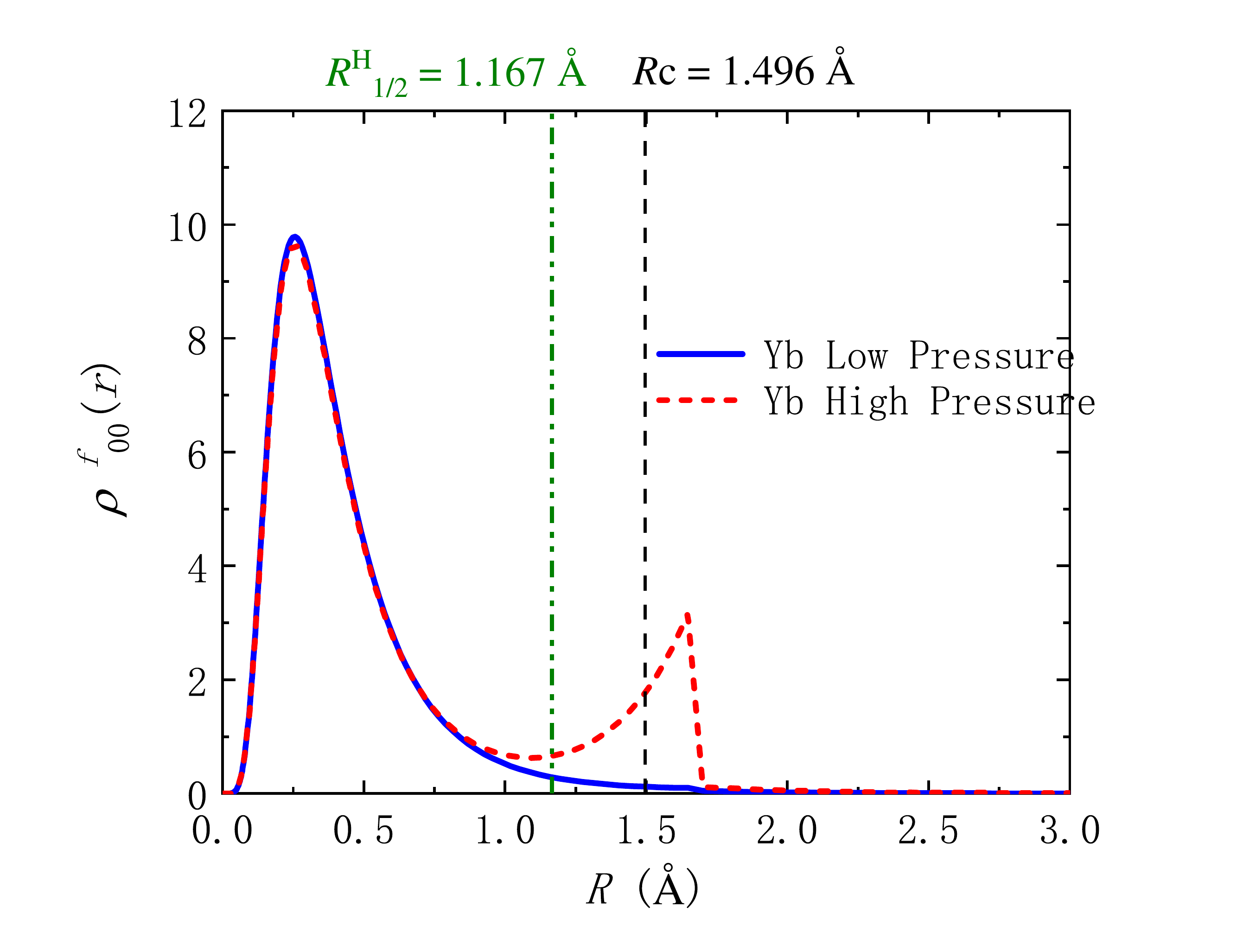}
\caption{\label{fig:fig6} The spherical averaged radial charge distribution of Yb at low pressure
(blue solid line) and high pressure (red dashed line). $R_{1/2}^H$ is half of the distance between
the closest atoms in high pressure. $R_{\rm c}$ stands for the augmented range of PAW.}
\end{figure}

\begin{figure}[htbp]
\centering
\includegraphics[width=0.48\textwidth]{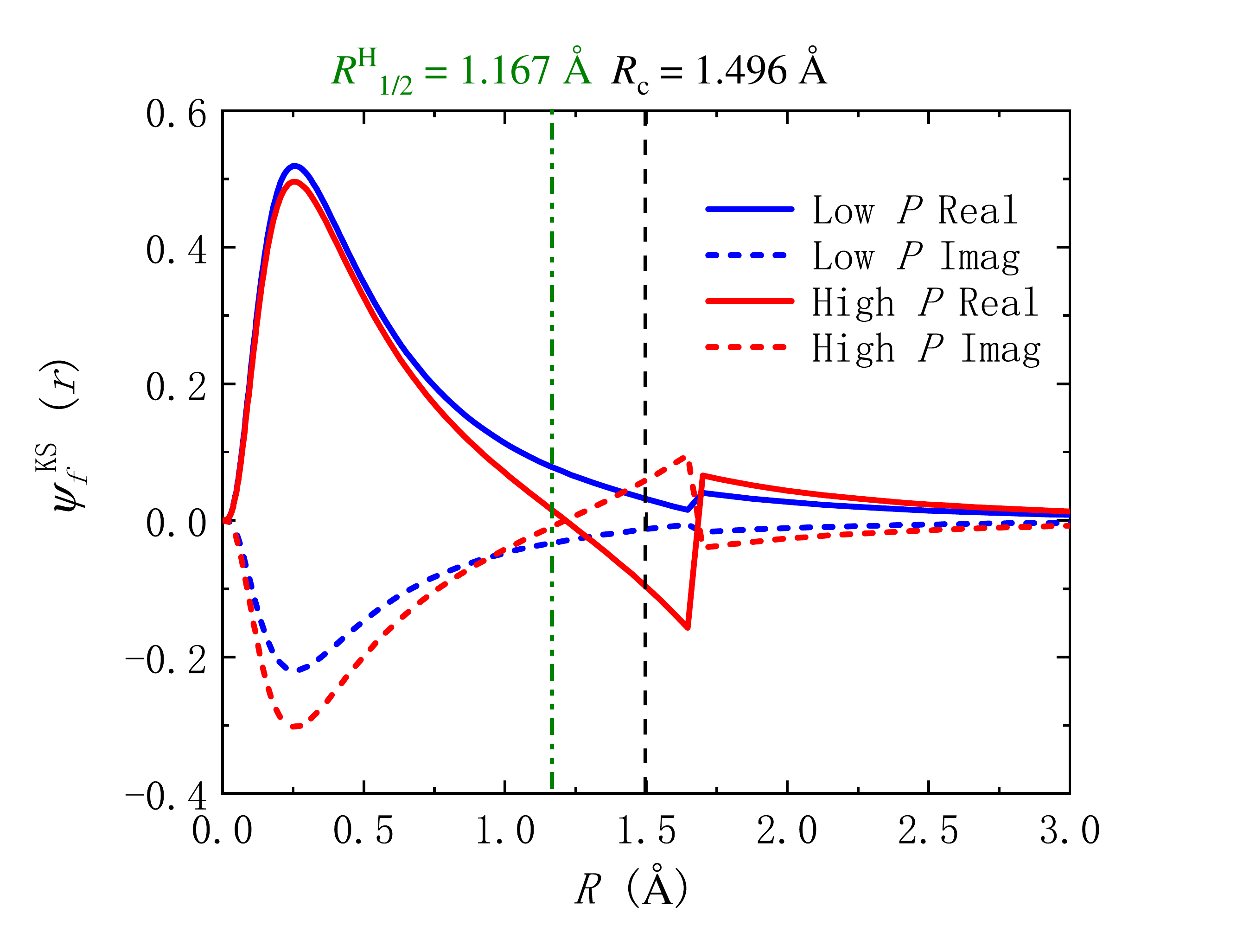}
\caption{\label{fig:fig7}The radial part of a KS orbital with f feature of Yb at low pressure
(blue lines) and high pressure (red lines). The solid line and the dashed line stand for real and
imaginary part of the orbital respectively. $R_{1/2}^H$ is half of the distance between the closest
atoms in high pressure. $R_{\rm c}$ stands for the augmented range of PAW.}
\end{figure}

In Fig.\ref{fig:fig6}, the average of radial charge distribution of 4$f$ orbital in Yb over a
spherical surface is shown at low (blue solid line) and high pressure (red dashed line). The
$R_{1/2}^H$ is the half of the distance between the nearby atoms at high pressure, which stands
for the radial not overlap. The radial charge can reflect the shape of the local orbital.
It can be observed that when R is smaller than $R_{1/2}^H$, the charge distributions of low and
high pressure are both similar to the shape of $F_1$. When in the range between $R_{1/2}^H$ and
$R_{\rm c}$, the high-pressure result shows are divergent feature like $F_2$. The divergent $F_2$
feature leading to additional electrons in the range between $R_{1/2}^H$ and $R_{\rm c}$. The charge
distribution in high-pressure with $F_2$ feature implies the local orbital lost the typical
characteristic of 4$f$ like $F_1$.

The Fig.\ref{fig:fig7} shows two radial parts of local orbital functions of Yb at low and high
pressure respectively. These local orbitals are from the occupied KS-orbitals close to the Fermi
energy at Gamma point, which are almost f-feature. It can be clearly seen that at low pressure
(blue lines), the orbital shows a typical 4$f$-feature as $F_1$, and almost no $F_2$ feature
expresses. For the orbital at high pressure, an $F_2$ feature shows in the range $R_{1/2}^H$ and
$R_{\rm c}$, which is consistent with the charge distribution results. Another important feature
of the high-pressure orbital is that a ``node'' appears just at the position $R_{1/2}^H$. It
reveals a probable explanation for the occupation over 1.0. The overlap of the augmented sphere
leads to the invasion of the nearest orbital, and makes the feature far from typical 4$f$. Then,
the weight of $F_2$ becomes larger at the projection operator. The feature of 4$f$ near the nucleus
still needs to be described with $F_1$. At last, the excessive $F_2$ component together with the
$F_1$ leads to an over 1.0 occupation.

\subsection{Abnormal behavior of ``+$U$'' in PAW}

It is demonstrated that the overlap of local orbital in PAW can cause an abnormal local density
matrix, and one can explain the abnormal behavior of simulation results by PAW from this point of
view. For the E-V, P-V and B-V results, the LAPW results increase faster with the decreasing of the
volume than PAW. It is the results of total energy correction with FLL double counting term and
local density matrix difference between PAW and LAPW. There are two different conditions. One is
the occupation of every local orbital is lower than 0.5 (as in La). In this condition, the total
energy grows faster when the occupation of the local density matrix grows faster, and the P-V and
B-V become steeper at a small volume. The other condition is that all local orbitals are over 0.5
occupations (as in Yb). At this condition, the total energy grows slower when the occupation of the
local density matrix grows faster, and the P-V, B-V becomes less steep at a small volume. The case
between these two conditions (as in Gd) leads to the cross of E-V, P-V and B-V curves between PAW
and LAPW results. Since the DFT results of PAW and LAPW are similar at both low and high-pressure
conditions, the contribution from the DFT part to the total energy in DFT+$U$ is similar. The main
difference between the DFT+$U$ results of PAW and LAPW comes from the ``+$U$'' correction.
The ``+$U$'' correction based on PAW in VASP has a larger projection range, thus more electrons
are counted in the local density matrix. Together with the overlap of local orbital, the local
orbital occupation increases faster in PAW than it in the LAPW with the decreasing of volume.
It makes the ``+$U$'' correction shows a more obvious effect in PAW results, which may even lead
to an unphysical result.

For the abnormal behavior of Yb in PAW results, that the B-V results show a decrease with the
volume decreasing, it is caused by a similar mechanism as above. In the high press condition of
Yb, the occupation of local orbitals is all over 1.0 in PAW. As having been introduced at the
beginning of this part, the contribution of the ``+$U$'' correction of energy becomes negative
when the occupation is over 1.0.
\textcolor{black}{With the volume decreasing and the occupation of 4$f$ increasing, the negative
contribution to the total energy in Yb becomes remarkable.}
This negative contribution
from the ``+$U$'' correction suppresses a normal trend described by the DFT part. The P-V should
also show a similar behavior in the PAW, if the volume is smaller.

The difference of electron structure results between PAW and LAPW can be explained from two
aspects. Firstly, the large local orbital range in PAW has more weight in the KS orbital than LAPW
dose. And it makes ``+$U$'' correction in PAW has more influence on the band structure. Secondly,
the large projection range in PAW makes more occupation on the local orbitals, and more occupation
means lower orbital energy in ``+$U$'' correction. It leads to a lower band energy in occupied
orbitals in PAW. Though the electron structure results of PAW and LAPW show some differences, no
obvious qualitative difference as in the E-V, P-V and B-V results has been captured. It may be
understood from the fact that the ``+$U$'' correction of orbital is linear, but the correction of
total energy is quadratic.

\subsection{Underestimation of pressure in correction methods}
\textcolor{black}{Another phenomenon observed above is that the DFT predicts higher pressure than
other correction methods in small volume (high-pressure) range for Tm, Yb and Lu.
This underestimating of pressure appears in all correction methods we tested.
Similar results are reported in many ``$+U$'' and hybrid functional works at ambient pressure,
where the correction methods give a larger volume
and a smaller bulk modulus comparing with the DFT results\cite{Yu2019,Loschen2007}.}

\begin{figure}[htbp]
\centering
\includegraphics[width=0.48\textwidth]{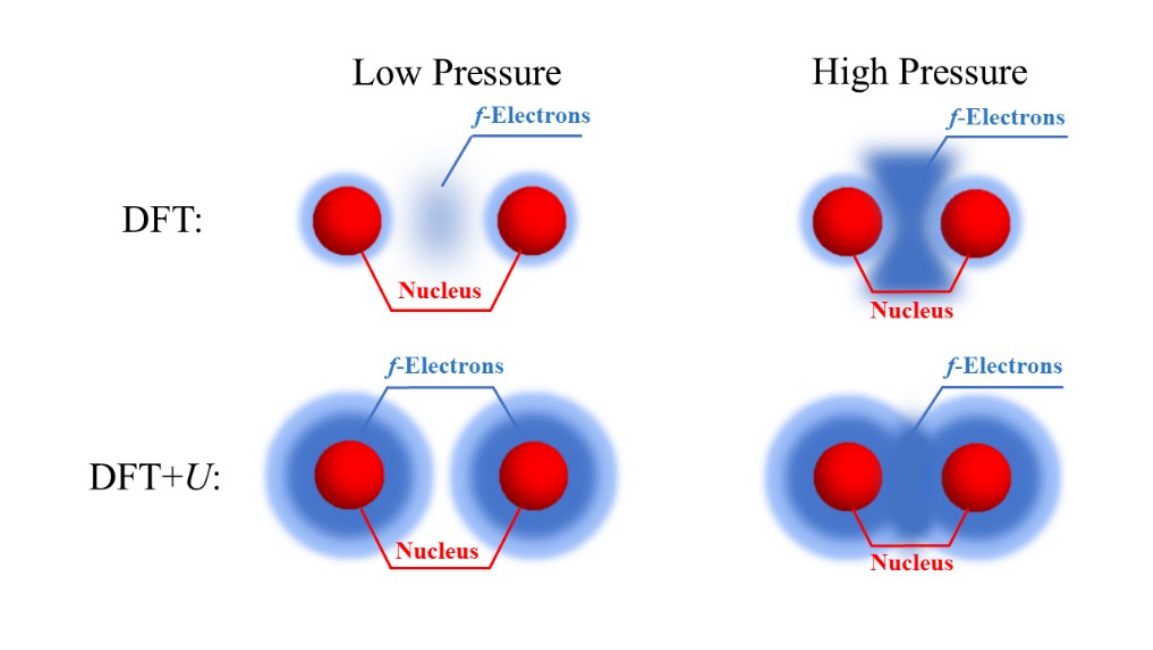}
\caption{\label{fig:ill}\textcolor{black}{The diagram about the underestimation of
pressure in correlation correction comparing with the common DFT.
The upper panel shows that in DFT, the electron density increases uniformly when the pressure becomes higher. The lower panel shows the correction methods (as DFT+$U$) stabilize the electron in the interatomic region in the high-pressure condition, which leads to a binding trend between atoms.
}}
\end{figure}

\textcolor{black}{A usual interpretation for this phenomenon is that the correction methods make the correlation
electrons more concentrated around the nuclei which leads the atoms behaving like isolated atoms.
The isolated behavior causes the binding between the atoms are weakened, and induce a larger volume
and a smaller bulk modulus. However, in high-pressure condition with hundreds of gigapascals, the
isolated atom interpretation may failed. We propose a qualitative understanding for the phenomenon
in high-pressure range from the point that the capability of localizing electrons in HSE06 and ``+$U$''
may increase the binding between atoms. Different from the condition at ambient pressure,
when the volume is small enough, there is not much space for the $f$ electrons concentrated
around the atoms. The electrons may prefer to localize between atoms, and the correction methods still
can stabilize these local electrons. The concentration of electrons between atoms can binding the nucleus
tight, which seems like some kind of ``bonding effect''. This effect makes the correction methods to
predict a much smaller volume in high-pressure condition, or smaller pressure with the same volume.
An illustration try to explain the picture is shown in Fig.\ref{fig:ill} schematically.
However, it is just a rough interpretation qualitatively, and more detailed investigation is
required for clearly understanding the results.
}

\section{Conclusion}
In summary, we investigate the performance of ``+$U$'' correction in high pressure based on the
widely-used augmented methods (PAW and LAPW) systematically. Our work exhibits two previously
unnoticed but vital phenomena in the high-pressure simulation of DFT+$U$, and highlights the
necessity of checking the applicability for DFT+$U$ simulation based on PAW for high-pressure
conditions.

On the one hand, we found that the force can be seriously underestimated in high-pressure
(over 100 GPa) with DFT+$U$ in VASP based on the PAW method. The underestimation can be even up to
50$\%$ comparing with the pressure calculated from DFT+$U$ in WIEN2k based on LAPW and HSE06 based
on PAW. Through delicate analysis, we point out that the larger range of projection operator
usually used in PAW of VASP than that in LAPW of WIEN2k should be the main cause of the different
behavior of ``+$U$'' correction in these two methods. The large augmented range in PAW makes the
overlap of local orbitals. And the overlap leads to an occupation more than 1.0 in the high-pressure
condition. From our investigation, it is caused by the normalization problem of local orbital in
small volume, which implies the over occupation is unphysical. The over occupation makes the
4$f$-local orbitals of PAW in VASP lose the radial feature of 4$f$, and weakens the reliability of
the DFT+$U$ method in high-pressure conditions. Thus, we propose that the ``+$U$'' correction should
be carefully checked before applying to the simulation of high-pressure condition.

On the other hand, an unusual behavior of the pressure simulation between DFT and correlation
correction methods (DFT+$U$ and HSE06) is observed at high pressure. The pressure predicted from
correlation correction methods is unexpectedly lower than that from DFT at the same volume for some
heavy lanthanides (Tm, Yb and Lu). We explained the phenomenon from the aspect that the electron in interatomic region is stabilized in correction methods comparing with DFT. However, it is just a preliminary
explanation, and more detailed investigations are required for a more clearly understanding.
For further investigation, researchers should put efforts on the solvation of the
unphysical behavior of PAW in high pressure and the profound understanding of the
softening in correction methods. There are still some other factors not considered in our work,
such as spin polarization, spin-orbit coupling, phase transition and so on. Their influence should
be studied in the following researches.

\section{Acknowledgement}
We thank Yuan-Ji Xu, Hua-Jie Chen, Bo Sun, He-Guang Mao, Xing-Jie Han and
Dan Jian for helpful discussions. The work was supported by the National Nature Science Foundation of China (No.U1930401, NO.12004048 and No.11971066), the Science Challenge Project (Grant No.TZ2018002), the National Key R\&D Program of China 2021YFB3501503 and the Foundation of LCP. We thank the Tianhe platforms at the National Supercomputer Center in Tianjin.

\section{Appendix}

\subsection{Projection operator in LAPW}

In the frame of LAPW, the projection operator acts on the single electron orbital
$\psi_i(\mathbf{r})$ to obtain local density matrix, which has the formation like,

\begin{align}
n_{mm^\prime}=\sum_{i\in occ. }\int{d\mathbf{r}d\mathbf{r}^
\prime{\psi_i^\ast\left(\mathbf{r}\right)\hat{P}}_{mm^\prime}\left(\mathbf{r},
\mathbf{r}^\prime\right)\psi_i(\mathbf{r}^\prime)}
\end{align}
$n_{mm^\prime}$ is the element of matrix with quantum number $m$ and $m'$.
The summation over all electron orbitals $i$, which are occupied.
In LAPW methods, the radial part of the projection operator is determined by the radial part
of the augmented wave function and is truncated by the radius of augmented range.
Thus ${\hat{P}}_{mm^\prime}\left(\mathbf{r},\mathbf{r}^\prime\right)$ only extract the
information of subspace by the angular part. The ${\hat{P}}_{mm^\prime}\left(\mathbf{r},
\mathbf{r}^\prime\right)$ can be written as,
\begin{align}
{\hat{P}}_{mm^\prime}\left(\mathbf{r},\mathbf{r}^\prime\right)=\frac{1}{r^2}\theta_\alpha
\left(r\right)\delta\left(r-r^\prime\right)Y_{lm^\prime}^\ast\left({\hat{\mathbf{r}}}^\prime\right)
Y_{lm}(\hat{\mathbf{r}})
\end{align}
$\theta_\alpha\left(r\right)$ is a step function which is 1 within the augmented range
$\alpha$, and is 0 outside. $\delta\left(r-r^\prime\right)$ is a delta function.
$Y_{lm}(\hat{\mathbf{r}})$ is the spherical harmonic with angular momentum $l,\ m$.
Since ${\hat{P}}_{mm^\prime}\left(\mathbf{r},\mathbf{r}^\prime\right)$ do not specify
the feature of the radial part of local orbital, it is important to show the form of radial
wave function if LAPW orbitals. Here shows a basis with reciprocal coordinate
$\mathbf{k}+\mathbf{G}$
\begin{align}
\nonumber
&\phi_{\mathbf{k}+\mathbf{G}}\left(\mathbf{r}\right) =
\theta_I\left(\mathbf{r}\right)\Omega^{-\frac{1}{2}}\exp{\left(\mathrm{i}\left(\mathbf{k} +
\mathbf{G}\right)\cdot\mathbf{r}\right)}
\\ &+\sum_{\alpha lm}{\theta_\alpha\left(\mathbf{r}\right)\left[A_{\alpha lm}^{\mathbf{k}+\mathbf{G}}
u_{\alpha l}\left(r_\alpha\right)+B_{\alpha lm}^{\mathbf{k}+\mathbf{G}}{\dot{u}}_{\alpha l}
\left(r_\alpha\right)\right]Y_{lm}({\hat{\mathbf{r}}}_\alpha)}
\end{align}
$\alpha$ and $I$ stand for the augmented and interval ranges respectively.
$\theta_I\left(\mathbf{r}\right)$ and $\theta_\alpha\left(\mathbf{r}\right)$ are the step function,
which makes corresponding restricted in the certain range. In the interval range, the plane wave is
used. In augmented spherical range an atom-like wave function is used. $u_{\alpha l}\left(r_\alpha
\right)$ and ${\dot{u}}_{\alpha l}\left(r_\alpha\right)$ are the solution of radial Schrodinger
function at certain energy with spherical averaged KS potential and its derivative function to
energy respectively. The KS single electron orbital can be described by a set of these basis.
\begin{align}
\psi_i \left(\mathbf{r}\right)=\sum_{\mathbf{G}}{C_\mathbf{G}^{i}\phi_{i,\mathbf{G}}(\mathbf{r})}
\end{align}
The summation is over reciprocal lattice $\mathbf{G}$. $i$ stands for the order of KS orbital.
Thus, in the augmented range, the KS orbital is
\begin{align}
\psi_i \left(\mathbf{r}\right)=\sum_{lm}{\left[\mathcal{A}_{\alpha lm}^{i}u_{\alpha l}
\left(r\right)+\mathcal{B}_{\alpha lm}^{i}{\dot{u}}_{\alpha l}\left(r\right)\right]Y_{lm}
(\hat{\mathbf{r}})}
\end{align}
with $\mathcal{A}_{\alpha lm}^{i}=\sum_{\mathbf{G}}{C_\mathbf{G}^{i}A_{\alpha lm}^
{i,\mathbf{G}}}$ and $\mathcal{B}_{\alpha lm}^{i}=\sum_{\mathbf{G}}{C_\mathbf{G}^{i}B_{\alpha lm}^
{i,\mathbf{G}}}$.

Since $\left\langle u_i\middle| u_i\right\rangle=1$ and
$\left\langle u_i\middle|{\dot{u}}_i\right\rangle=0$ in LAPW method,
the element of local density matrix is
\begin{align}
n_{mm^\prime} =\sum_{i\in occ.}{\mathcal{A}_{\alpha lm}^{i}\left[\mathcal{A}_
{\alpha l m^\prime}^{i}\right]^\ast+\mathcal{B}_{\alpha lm}^{i}\left[\mathcal{B}_
{\alpha l m^\prime}^{i}\right]^\ast\left\langle{\dot{u}}_l\middle|{\dot{u}}_l\right\rangle}
\end{align}

\subsection{Projection operator in PAW}

PAW is another important augmented method.
The formation of the local density matrix and projection operator is the same as the ones in
LAPW methods. The difference of the element in the local density matrix between PAW and LAPW comes
from the different radial orbital of these two methods. In PAW, the KS orbital has the form,
\begin{align}
\left|\psi_i \right\rangle=\left|{\widetilde{\psi}}_i \right\rangle+\sum_{s}
{\left(\left|\phi_s\right\rangle-\left|{\widetilde{\phi}}_s\right\rangle\right)
\langle{\widetilde{p}}_s|{\widetilde{\psi}}_i\rangle}
\end{align}
$\left|\phi_s\right\rangle$ and $\left|{\widetilde{\phi}}_s\right\rangle$ are also atom-like
orbitals with the form,
\begin{align}
\phi_s\left(\mathbf{r}\right)=R_s\left(r\right)Y_s(\hat{\mathbf{r}})
\\
{\widetilde{\phi}}_s\left(\mathbf{r}\right)={\widetilde{R}}_s\left(r\right)Y_s(\hat{\mathbf{r}})
\end{align}
Subscript $s$ is a collection of angular momentum $lm$, atom site
$\alpha$ and the order of radial function $t$. With the definition, the element of
local density matrix in PAW is
\begin{align}
\label{eq:pawnn}
\nonumber
n_{mm^\prime} &=\sum_{i\in occ.}{\left\langle\psi_i \right|{\hat{P}}_{mm^\prime}
\left|\psi_i \right\rangle}
\\
\nonumber
&=\sum_{i\in occ.}{\sum_{t_1t_2}\left\langle{\widetilde{\psi}}_i \middle|
{\widetilde{p}}_{mt_1}\right\rangle\left\langle\phi_{mt_1}\right|\mathbf{P}_{mm^\prime}
\left|\phi_{m^\prime t_2}\right\rangle\left\langle{\widetilde{p}}_{m^\prime t_2}\middle|
{\widetilde{\psi}}_{i}\right\rangle}
\\
&=\sum_{t_1t_2}{\rho_{mm^\prime}^{t_1t_2}\left\langle R_{t_1}\middle| R_{t_2}\right\rangle}
\end{align}
and
\begin{align}
\rho_{mm^\prime}^{t_1t_2}=\sum_{i\in occ.}{\left\langle{\widetilde{\psi}}_i \middle|
{\widetilde{p}}_{{mt}_1}\right\rangle\left\langle{\widetilde{p}}_{m^\prime t_2}\middle|
{\widetilde{\psi}}_i \right\rangle}
\end{align}

\clearpage


\begin{thebibliography}{10}

\bibitem{Bushman1992}
A. V. Bushman.
\newblock Intense Dynamic Loading of Condensed Matter (CRC Press, 1992).

\bibitem{Feynman1949}
R. P. Feynman, N. Metropolis, and E. Teller.
\newblock Equations of State of Elements Based on the Generalized Fermi-Thomas Theory.
\newblock {\em Phys. Rev.}, 75, 1561, 1949.

\bibitem{McCarthy1965}
S. L. McCarthy.
\newblock The Kirzhnits Corrections to the Thomas-Fermi Equation of State, 1965.

\bibitem{Kirzhnits1975}
D. A. Kirzhnits, Y. E. Lozovik, and G. V. Shpatakovskaya.
\newblock Statistical Model of Matter.
\newblock {\em Sov. Phys. Uspekhi} 18, 649, 1975.

\bibitem{Liu2016}
H. Liu, H. Song, Q. Zhang, G. Zhang, and Y. Zhao.
\newblock Validation for Equation of State in Wide Regime: Copper as Prototype,
\newblock {\em Matter Radiat. Extrem.} 1, 123, 2016.

\bibitem{Young2016}
D. A. Young, H. Cynn, P. S\"{o}derlind, and A. Landa.
\newblock Zero-Kelvin Compression Isotherms of the Elements 1$\leq$ Z$\leq$ 92 to 100 GPa.
\newblock {\em J. Phys. Chem. Ref. Data} 45, 043101, 2016.

\bibitem{Song2009}
H.-F. Song, H.-F. Liu, G.-C.Zhang, Y.-H.Zhao.
\newblock Numerical Simulation of Wave Propagation and
Phase Transition of Tin under Shock-Wave Loading.
\newblock {\em Chin. Phys. Lett.} 26, 066401, 2009.

\bibitem{Song2007}
H.-F. Song, H.-F. Liu.
\newblock Theoretical Study of Thermodynamic Properties of Metal Be.
\newblock {\em Acta Phys. Sin.} 5, 2007.

\bibitem{Grasso2013}
V. B. Grasso.
\newblock Rare Earth Elements in National Defense:
Background, Oversight Issues, and Options for Congress, in
(Library Of Congress Washington DC Congressional Research Service, 2013).

\bibitem{Samudrala2013}
G. K. Samudrala and Y. K. Vohra.
\newblock Structural Properties of Lanthanides at Ultra High Pressure,
in Handbook on the Physics and Chemistry of Rare Earths, Vol. 43 (Elsevier, 2013), pp. 275--319.

\bibitem{Finnegan2021}
S. E. Finnegan, C. V. Storm, E. J. Pace, M. I. McMahon, S. G. MacLeod,
E. Plekhanov, N. Bonini, and C. Weber.
\newblock High-Pressure Structural Systematics in Neodymium up to 302 GPa.
\newblock {\em Phys. Rev. B} 103, 134117, 2021.

\bibitem{Perreault2020}
C. S. Perreault and Y. K. Vohra.
\newblock Static Compression of Rare Earth Metal Holmium to 282 GPa.
\newblock {\em High Press. Res.} 40, 392, 2020.

\bibitem{Sun2016}
H. Sun, D. Kang, J. Dai, W. Ma, L. Zhou, and J. Zeng.
\newblock First-Principles Study on Equation of States
and Electronic Structures of Shock Compressed Ar up to Warm Dense Regime.
\newblock {\em J. Chem. Phys.} 144, 124503, 2016.

\bibitem{Tsuchiya2003}
T. Tsuchiya.
\newblock First-Principles Prediction of the P-V-T Equation of State
of Gold and the 660-Km Discontinuity in Earth's Mantle:
First-Principles Prediction of Equation of State of Gold.
\newblock {\em J. Geophys. Res. Solid Earth.} 108, 2003.

\bibitem{Zhang2017}
S. Zhang, K. P. Driver, F. Soubiran, and B. Militzer.
\newblock Equation of State and Shock Compression of Warm Dense Sodium--A First-Principles Study.
\newblock {\em J. Chem. Phys.} 146, 074505, 2017.

\bibitem{Martin2004}
R. M. Martin.
\newblock Electronic Structure: Basic Theory and Practical Methods
(Cambridge University Press, Cambridge, UK; New York, 2004).

\bibitem{Jain2016}
A. Jain, Y. Shin, and K. A. Persson.
\newblock Computational Predictions of Energy Materials Using Density Functional Theory.
\newblock {\em Nat. Rev. Mater.} 1, 15004, 2016.

\bibitem{Soderlind2018}
P. S{\"o}derlind and D. Young.
\newblock Assessing Density-Functional Theory for Equation-Of-State.
\newblock {\em Computation} 6, 13, 2018.

\bibitem{Zhang2021}
T. Zhang, Y. Wang, J. Xian, S. Wang, J. Fang, S. Duan, X. Gao, H. Song, and H. Liu.
\newblock Effect of the Projector Augmented Wave
Potentials on the Simulation of Thermodynamic Properties of Vanadium.
\newblock {\em Matter Radiat. Extrem.} 6, 068401, 2021.

\bibitem{Weinert1981}
M. Weinert.
\newblock Solution of Poisson's Equation: Beyond Ewald-type Methods.
\newblock {\em J. Math. Phys.} 22, 2433, 1981.

\bibitem{Wimmer1981}
E. Wimmer, H. Krakauer, M. Weinert, and A. J. Freeman.
\newblock Full-Potential Self-Consistent Linearized-Augmented-Plane-Wave
Method for Calculating the Electronic Structure of Molecules and Surfaces:
$\rm O_2$ Molecule, \newblock {\em Phys. Rev. B} 24, 864, 1981.

\bibitem{Vanderbilt1990}
D. Vanderbilt.
\newblock Soft Self-Consistent Pseudopotentials in a Generalized Eigenvalue Formalism.
\newblock {\em Phys. Rev. B} 41, 7892, 1990.

\bibitem{Blochl1994}
P. E. Bl{\"o}chl.
\newblock Projector Augmented-Wave Method. \newblock {\em Phys. Rev. B} 50, 17953, 1994.

\bibitem{Blaha2020}
P. Blaha, K. Schwarz, F. Tran, R. Laskowski, G. K. H. Madsen, and L. D. Marks.
\newblock WIEN2k: An APW+lo Program for Calculating the Properties of Solids.
\newblock {\em J. Chem. Phys.} 152, 074101, 2020.

\bibitem{Elk}
J. K. Dewhurst et al.
\newblock The Elk Code, elk.sourceforge.net.

\bibitem{Kresse1999}
G. Kresse and D. Joubert.
\newblock From Ultrasoft Pseudopotentials to the Projector Augmented-Wave Method.
\newblock {\em Phys. Rev. B} 59, 1758, 1999.

\bibitem{Giannozzi2020}
P. Giannozzi, O. Baseggio, P. Bonf\`a, D. Brunato, R. Car, I. Carnimeo,
C. Cavazzoni, S. de Gironcoli, P. Delugas, F. Ferrari Ruffino,
A. Ferretti, N. Marzari, I. Timrov, A. Urru, and S. Baroni.
\newblock QUANTUM ESPRESSO toward the Exascale.
\newblock {\em J. Chem. Phys.} 152, 154105, 2020.

\bibitem{Enkovaara2010}
J. Enkovaara, C. Rostgaard, J. J. Mortensen, J. Chen, M. Du\l{}ak,
L. Ferrighi, J. Gavnholt, C. Glinsvad, V. Haikola, H. A. Hansen,
H. H. Kristoffersen, M. Kuisma, A. H. Larsen, L. Lehtovaara, M. Ljungberg,
O. Lopez-Acevedo, P. G. Moses, J. Ojanen, T. Olsen, V. Petzold, N. A. Romero, J. Stausholm-M\o ller,
M. Strange, G. A. Tritsaris, M. Vanin, M. Walter, B. Hammer, H. H\"{a}kkinen,
G. K. H. Madsen, R. M. Nieminen, J. K. N\o rskov, M. Puska, T. T. Rantala, J. Schi\o tz,
K. S. Thygesen, and K. W. Jacobsen.
\newblock Electronic Structure Calculations with GPAW:
A Real-Space Implementation of the Projector Augmented-Wave Method.
\newblock {\em J. Phys. Condens. Matter} 22, 253202, 2010.

\bibitem{Gonze2020}
X. Gonze, B. Amadon, G. Antonius, F. Arnardi, L. Baguet, J.-M. Beuken, J. Bieder, F. Bottin, J. Bouchet, E. Bousquet, N. Brouwer, F. Bruneval, G. Brunin, T. Cavignac, J.-B. Charraud, W. Chen, M. Côté, S. Cottenier, J. Denier, G. Geneste, P. Ghosez, M. Giantomassi, Y. Gillet, O. Gingras, D. R. Hamann, G. Hautier, X. He, N. Helbig, N. Holzwarth, Y. Jia, F. Jollet, W. Lafargue-Dit-Hauret, K. Lejaeghere, M. A. L. Marques, A. Martin, C. Martins, H. P. C. Miranda, F. Naccarato, K. Persson, G. Petretto, V. Planes, Y. Pouillon, S. Prokhorenko, F. Ricci, G.-M. Rignanese, A. H. Romero, M. M. Schmitt, M. Torrent, M. J. van Setten, B. Van Troeye, M. J. Verstraete, G. Zérah, and J. W. Zwanziger.
\newblock The Abinitproject: Impact, Environment and Recent Developments.
\newblock {\em Comput. Phys. Commun.} 248, 107042, 2020.

\bibitem{Holzwarth1997}
N. A. W. Holzwarth, G. E. Matthews, R. B. Dunning, A. R. Tackett, and Y. Zeng.
\newblock Comparison of the Projector Augmented-Wave, Pseudopotential,
and Linearized Augmented-Plane-Wave Formalisms for Density-Functional Calculations of Solids.
\newblock {\em Phys. Rev. B} 55, 2005, 1997.

\bibitem{Shick1999}
A. B. Shick, A. I. Liechtenstein, and W. E. Pickett.
\newblock Implementation of the LDA+U Method Using the
Full-Potential Linearized Augmented Plane-Wave Basis.
\newblock {\em Phys. Rev. B} 60, 10763, 1999.

\bibitem{Anisimov1997}
V. I. Anisimov, F. Aryasetiawan, and A. I. Lichtenstein.
\newblock First-Principles Calculations of the Electronic Structure
and Spectra of Strongly Correlated Systems: The LDA + $U$ Method.
\newblock {\em J. Phys. Condens.} Matter 9, 767, 1997.

\bibitem{Liechtenstein1995}
A. I. Liechtenstein, V. I. Anisimov, and J. Zaanen.
\newblock Density-Functional Theory and Strong Interactions:
Orbital Ordering in Mott-Hubbard Insulators.
\newblock {\em Phys. Rev. B} 52, R5467, 1995.

\bibitem{Dompablo2011}
M. E. Arroyo-de Dompablo, A. Morales-Garc\'{i}a, and M. Taravillo.
\newblock DFT+ U Calculations of Crystal Lattice, Electronic Structure,
and Phase Stability under Pressure of ${\rm TiO_2}$ Polymorphs.
\newblock {\em J. Chem. Phys.} 135, 054503, 2011.

\bibitem{Himmetoglu2014}
B. Himmetoglu, A. Floris, S. de Gironcoli, and M. Cococcioni.
\newblock Hubbard-Corrected DFT Energy Functionals: The LDA+U Description of Correlated Systems.
\newblock {\em Int. J. Quantum Chem.} 114, 14, 2014.

\bibitem{Deng2008}
X. Deng, X. Dai, and Z. Fang.
\newblock LDA + Gutzwiller Method for Correlated Electron Systems.
\newblock {\em EPL Europhys. Lett.} 83, 37008, 2008.

\bibitem{Anisimov2010}
V. Anisimov and Y. Izyumov.
\newblock Electronic Structure of Strongly Correlated Materials
(Springer-Verlag, Berlin Heidelberg, 2010).

\bibitem{Avella2012}
A. Avella and F. Mancini.
\newblock Strongly Correlated Systems. Theoretical Methods
(Springer, Heidelberg; New York, 2012).

\bibitem{Casadei2016}
M. Casadei, X. Ren, P. Rinke, A. Rubio, and M. Scheffler.
\newblock Density Functional Theory Study of the $\alpha$ - $\gamma$
Phase Transition in Cerium: Role of Electron Correlation and $f$-Orbital Localization.
\newblock {\em Phys. Rev. B} 93, 075153, 2016.

\bibitem{Perdew1996}
J. P. Perdew, M. Ernzerhof, and K. Burke.
\newblock Rationale for Mixing Exact Exchange with Density Functional Approximations.
\newblock {\em J. Chem. Phys.} 105, 9982, 1996.

\bibitem{Jiang2015}
H. Jiang.
\newblock First-Principles Approaches for Strongly Correlated Materials:
A Theoretical Chemistry Perspective.
\newblock {\em Int. J. Quantum Chem.} 115, 722, 2015.

\bibitem{Rahm2009}
M. Rahm and N. V. Skorodumova.
\newblock Phase Stability of the Rare-Earth Sesquioxides under Pressure.
\newblock {\em Phys. Rev. B} 80, 104105, 2009.

\bibitem{Modak2013}
P. Modak and A. K. Verma.
\newblock First-Principles Investigations of Equation of States and Phase
Transitions in PaN under Pressure. in (Indian Institute of Technology, Bombay, Mumbai, India, 2013),
pp. 80--81.

\bibitem{Steneteg2012}
P. Steneteg, B. Alling, and I. A. Abrikosov.
\newblock Equation of State of Paramagnetic CrN from Ab Initio Molecular Dynamics.
\newblock {\em Phys. Rev. B} 85, 144404, 2012.

\bibitem{Hsu2011}
H. Hsu, K. Umemoto, M. Cococcioni, and R. M. Wentzcovitch.
\newblock The Hubbard U Correction for Iron-Bearing Minerals:
A Discussion Based on (Mg,Fe)${\rm SiO_3}$ Perovskite.
\newblock {\em Phys. Earth Planet. Inter.} 185, 13, 2011.

\bibitem{Rollmann2005}
G. Rollmann, P. Entel, A. Rohrbach, and J. Hafner.
\newblock High-Pressure Characteristics of ${\alpha-\rm Fe_{2}O_{3}}$ Using DFT+ U.
\newblock {\em Phase Transit.} 78, 251, 2005.

\bibitem{Geng2007}
H. Y. Geng, Y. Chen, Y. Kaneta, and M. Kinoshita.
\newblock Structural Behavior of Uranium Dioxide under Pressure by LSDA + U Calculations.
\newblock {\em Phys. Rev. B} 75, 054111, 2007.

\bibitem{Song2009-2}
Z. Song, J. J. Yang, and J. S. Tse.
\newblock A Comparative Study on the LDA + U and Hybrid Functional
Methods on the Description of the Electronic Structure of ${\rm YTiO_3}$ under High Pressure.
\newblock {\em Can. J. Chem.} 87, 1374, 2009.

\bibitem{Zhang2010}
P. Zhang, B.-T. Wang, and X.-G. Zhao.
\newblock Ground-State Properties and High-Pressure Behavior of
Plutonium Dioxide: Density Functional Theory Calculations.
\newblock {\em Phys. Rev. B} 82, 144110, 2010.

\bibitem{Wen2013}
X.-D. Wen, R. L. Martin, G. E. Scuseria, S. P. Rudin, E. R. Batista, and A. K. Burrell.
\newblock Screened Hybrid and DFT + U Studies of the Structural,
Electronic, and Optical Properties of ${\rm U_{3}O_{8}}$.
\newblock  {\em J. Phys. Condens. Matter} 25, 025501, 2013.

\bibitem{Zhou2014}
L. Zhou, F. K\"{o}rmann, D. Holec, M. Bartosik, B. Grabowski, J. Neugebauer, and P. H. Mayrhofer.
\newblock Structural Stability and Thermodynamics of CrN Magnetic Phases
from Ab Initio Calculations and Experiment
\newblock {\em Phys. Rev. B} 90, 184102, 2014.

\bibitem{Wang2016}
Y.-C. Wang, Z.-H. Chen, and H. Jiang.
\newblock The Local Projection in the Density Functional
Theory plus U Approach: A Critical Assessment.
\newblock {\em J. Chem. Phys.} 144, 144106, 2016.

\bibitem{Perdew1996-2}
J. P. Perdew, K. Burke, and M. Ernzerhof.
\newblock Generalized Gradient Approximation Made Simple.
\newblock {\em Phys. Rev. Lett.} 77, 3865, 1996.

\bibitem{Jiang2009}
H. Jiang, R. I. Gomez-Abal, P. Rinke, and M. Scheffler.
\newblock Localized and Itinerant States in Lanthanide Oxides United by GW @ LDA + U.
\newblock {\em Phys. Rev. Lett.} 102, 2009.

\bibitem{Moree2018}
J. B. Mor\'{e}e and B. Amadon.
\newblock First-Principles Calculation of Coulomb Interaction
Parameters for Lanthanides: Role of Self-Consistence and Screening Processes.
\newblock {\em Phys. Rev. B} 98, 205101, 2018.

\bibitem{Heyd2003}
J. Heyd, G. E. Scuseria, and M. Ernzerhof,
\newblock Hybrid Functionals Based on a Screened Coulomb Potential.
\newblock {\em J. Chem. Phys.} 118, 8207, 2003.

\bibitem{Ryee2018}
S. Ryee and M. J. Han.
\newblock The Effect of Double Counting, Spin Density,
and Hund Interaction in the Different DFT+U Functionals.
\newblock {\em Sci. Rep.} 8, 9559, 2018.

\bibitem{ORegan2010}
D. D. O'Regan, N. D. M. Hine, M. C. Payne, and A. A. Mostofi.
\newblock Linear-scaling DFT + $U$ with full local orbital optimization.
\newblock {\em Phys. Rev. B} 85, 085107, 2012.

\bibitem{Bengone2000}
O. Bengone, M. Alouani, P. Bl\"{o}chl, and J. Hugel.
\newblock Implementation of the Projector
Augmented-Wave LDA+U Method: Application to the Electronic Structure of NiO.
\newblock {\em Phys. Rev. B} 62, 16392, 2000.

\bibitem{Cococcioni2005}
M. Cococcioni and S. de Gironcoli.
\newblock Linear Response Approach to
the Calculation of the Effective Interaction Parameters in the LDA + U Method.
\newblock {\em Phys. Rev. B} 71, 2005.

\bibitem{Ylvisaker}
E. R. Ylvisaker, DFT and DMFT:
Implementations and Applications to the Study of Correlated Materials, 178 (n.d.).

\bibitem{Yu2019}
J. Yu, M. Zhang, Z. Zhang, S. Wang and Y. Wu,
\newblock Hybrid-functional calculations of electronic structure and phase stability of MO (M = Zn{,} Cd{,} Be{,} Mg{,} Ca{,} Sr{,} Ba) and related ternary alloy M$_{\rm x}$Zn$_{\rm {1-x}}$O
\newblock {\em RSC Adv.} 9, 2019.

\bibitem{Loschen2007}
C. Loschen, J. Carrasco, K. M. Neyman and F. Illas,
\newblock First-principles $\mathrm{LDA}+\mathrm{U}$ and $\mathrm{GGA}+\mathrm{U}$ study of cerium oxides: Dependence on the effective U parameter
\newblock {\em Phys. Rev. B} 75, 2007.

\end{thebibliography}
\end{document}